\documentclass[pre,pre,twocolumn,floatfix,showpacs]{revtex4}
\usepackage{graphicx}
\usepackage{epsfig}
\usepackage{amssymb,amsmath}
\usepackage{bm}

\def\3dots{\:\raisebox{-0.5ex}{$\stackrel{\textstyle.}{:}$}\:}
\def\beq{\begin{equation}}
\def\eeq{\end{equation}}
\def\bea{\begin{eqnarray}}
\def\eea{\end{eqnarray}}

\begin{document}
\title{Poisson's ratio in composite elastic media with rigid rods}
\author{Moumita Das}
\author{F.C.\ MacKintosh}
\affiliation{Department of Physics and Astronomy, Vrije Universiteit, Amsterdam, The Netherlands. \\
 }

\begin{abstract}
We calculate both the micro-mechanical response and bulk elastic constants of composites of rods embedded in elastic media. We find two fixed points for Poisson's ratio with respect to rod density: there is an unstable fixed point for Poisson's ratio=1/2 (an incompressible system) and a stable fixed point for Poisson's ratio=1/4 (a compressible system). We also derive approximate expressions for the elastic constants for arbitrary rod density, which agree with exact results for both low and high density. These results may help to explain recent experiments [Physical Review Letters 102, 188303 (2009)] that reported compressibility for composites of microtubules in F-actin networks.
\end{abstract}
\date{\today}
\maketitle

There are many natural examples of composite materials, combining multiple components with distinct elastic properties. Wood, bone and various tissues are all made of composites, as is each living cell \cite{alberts}. Many composites consist of a reinforcing constituent such as stiff fibers embedded in a weaker, less stiff matrix \cite{nanotube1}. By varying the relative concentration of the constituents, one can tune the elastic moduli, resulting in materials with remarkable properties \cite{Thorpe,Nanotube2}. The collective properties are more than merely the sum total of those of its constituents. Among the most prevalent natural or living composites is the cell cytoskeleton, which consists of a complex scaffold of several distinct filamentous proteins, some of which are very rigid. Most previous biophysical studies of cytoskeletal networks have focused on purified gels or networks consisting of one type of filament \cite{hinner,tharmann,liu,kees,head,wilhelm,emt,Onck,NatMat,Koenderink}. The cytoskeleton, however, contains three major types of filaments: microtubules (MTs), filamentous actin (F-actin), and intermediate filaments. These filaments have vastly different bending stiffness. A few studies of reconstituted composite cytoskeletal networks have shown viscoelastic properties distinct from single-component networks \cite{gardelfilamin,karenpre,chaseprl}. A very recent experimental report also provides evidence of anomalous compressibility with the addition of stiff microtubules to a soft matrix \cite{kilfoil}.

Here, we develop a model for the mechanical response of a composite material consisting of rods in an elastic matrix, using a mean field approach and a dipole approximation for the rod-like inclusions. The elastic matrix under consideration is treated as an effective medium that is made of the bare elastic medium (e.g., the F-actin matrix) and a collection of rods (MTs) embedded in it. Consistent with the experiments of Ref.~\cite{kilfoil}, we find that the addition of rigid rods can lead to enhanced compressibility of an initially nearly incompressible medium. Specifically, we find that for matrices characterized by Poisson's ratio $1/4<\nu<1/2$, the addition of rods reduces $\nu$, while for $\nu<1/4$, stiff rods increase $\nu$. In this way, $\nu=1/4$ can be thought of as a stable fixed-point of such a composite. We further evaluate the Poisson's ratio and elastic moduli as functions of the concentration of rod-like inclusions. While this mean-field approach is only approximate at intermediate concentrations, we obtain an exact result in the limit of high concentration. 

\begin{figure}[t]
\includegraphics[width=8cm]{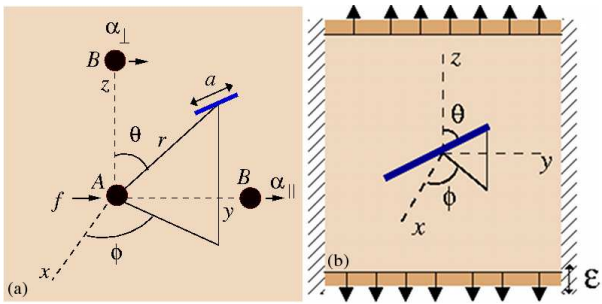}
\caption{\label{Fig1} (Color Online) We consider a point-force $f\hat y$ applied at the origin ($A$) and calculate the response at points $B$ located parallel (here, along the $y-$axis) and perpendicular (along the $z-$axis) to the applied force in Fig.~1 (a). The rod center of mass is located at a distance $r$ from the origin, and makes a polar angle $\theta$  and azimuthal angle $\phi$. The rod is oriented in a direction as shown in Fig.~1 (b).}
\end{figure}

We first study the micro-mechanics of our system using the elastic response function or Green's function for the displacement field $\vec u$ in response to an applied force. For an isotropic and homogenous elastic material with Lam\'e coefficients $\lambda$ and $\mu$, we can describe the displacement field $u_i$ at a position $\vec r$ in the medium due to a force $\vec f$ acting at point $\vec{r'}$, 
$
u_i(\vec{r})=\alpha_{ij} (\vec{r}-\vec{r'}) f_j (\vec{r'}),
$
with the linear response function $\alpha_{ij}$  given by :
\begin{equation}
 \alpha_{ij}(\vec{r})= \frac{1}{8 \pi \mu r} \big[ \hat{r}_i\hat{r}_j  (1- \beta)
+  \delta_{ij} (1 + \beta)\big],
\end{equation}
where $\beta = \mu/(\lambda + 2 \mu)$ is the ratio of the shear modulus $\mu$ to the longitudinal modulus. For an isotropic and homogeneous elastic material, $\alpha_{ij}(\vec r)$ reduces to just two distinct components corresponding to the response parallel and perpendicular to $\vec r$, as shown in Fig.\ \ref{Fig1}a. For an incompressible material in 3d, $\beta=0$ and the parallel and perpendicular response functions are related by a simple factor of two: $\alpha_{\parallel}(r)=2 \alpha_{\perp}(r)=1/4 \pi\mu r$, which is the elastic analogue of the Oseen Tensor \cite{landau,alex}. For compressible elastic media in 3d, $\beta>0$, $\alpha_\parallel (r)=1/4\pi\mu r$ and $\alpha_\perp (r)=(1+\beta)/8\pi\mu r$. Thus, $\beta$ provides a measure of the degree of compressibility. 

We calculate the change in the response function and Lam\'e coefficients upon addition of rods as follows. We consider a single rod of length $a$ embedded in the elastic medium, as shown in Fig.\ \ref{Fig1} (a). The presence of the rod represents a constraint on the displacement field induced by the applied force. For a force $f \hat{y}$ applied at $A$, the net displacement of the ends of the rod of length $a$ oriented as in Fig.~\ref{Fig1} (a) is given by $\Delta \vec u (\vec r)= (\vec u (\vec r+\vec a/2)-\vec u (\vec r-\vec a/2))$. We approximate the constraint of an incompressible rod by a dipole at its center of mass. This induced (tensile) dipole is oriented along the rod and its strength is chosen so as to enforce a constant end-to-end distance of the rod: $p= \mu \pi a (\vec a\cdot\Delta\vec u)/2=\chi\epsilon$, where $\chi=\mu\pi a^3/2$, $\epsilon=\hat a\cdot(\hat a \cdot \vec \nabla)\vec u$, and where we keep only leading-order terms in $a$, which is assumed to be small compared to the other lengths in the figure. The resulting displacement field at $B$ allows us to determine the change in the linear response functions. In our effective medium picture, we obtain the change in response as arising from a cloud of induced dipoles in the elastic continuum. 

\begin{figure}[t]
\includegraphics[width=8cm]{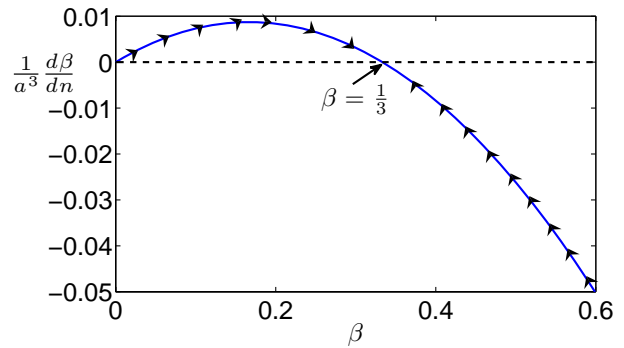}
\caption{\label{Fig2} (Color Online) The flow diagram for the degree of compressibility $\beta$ showing a stable fixed point  at $\beta = 1/3$, and an unstable fixed point at $\beta =0$.}
\end{figure}

\begin{figure}[b]
\includegraphics[width=8cm]{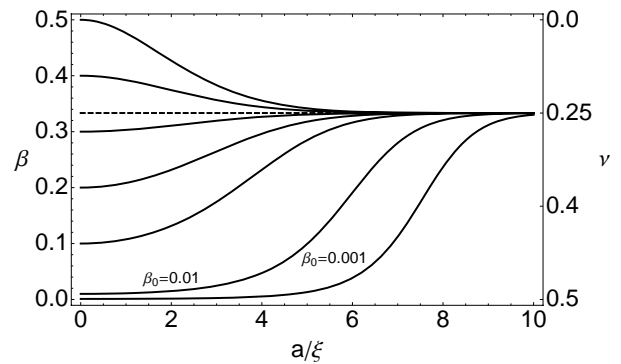}
\caption{\label{Fig3} The  degree of compressibility $\beta$ and Poisson's ratio $\nu$  of the composite as a function of mesh size $\xi$ for inextensible rods, for different values of the degree of compressibility $\beta_0$ of the medium in the absence of rods. The mesh size $\xi$ is related to the rod density $n$ by $1/\xi^{2}\equiv n a$.}
\end{figure}

The change in the parallel and perpendicular response functions with the addition of the rods are $\delta \alpha_{\parallel}=-(\pi/30)n a^3\alpha_\parallel$ and $\delta \alpha_{\perp} = \frac{1}{2}(1+3\beta^2)\delta\alpha_\parallel$, where $n$ is the rod number density and we have averaged over rod orientation. This is equivalent to $\delta \mu=\delta\lambda= \frac{1}{15} \chi n$. For a small increment $dn$ in added rods, we obtain the differential equation $d\beta/dn= \frac{\pi}{30}a^3\beta(1-3\beta)$. Thus, $\beta$ increases for $0<\beta<1/3$ and decreases for $\beta>1/3$, 
%
%
while $\beta$ is unchanged for $\beta=0$ and $\beta=1/3$. Therefore, for a slightly compressible medium, adding rods enhances the compressibility relative to the shear compliance, while for a highly compressible medium, the rods reduce the compressibility, as illustrated in Fig.~\ref{Fig2}.  For finite compressibility, addition of rods tends to drive the system towards a state with $\beta=1/3$. This suggests a stable fixed point (to the addition of rods) at $\beta=1/3$ and hence Poisson's ratio $\nu=1/4$. Similarly, $\beta=0$ and $\nu=1/2$ corresponds to an unstable fixed point. These results are also apparent from the full solution to the differential equation above for $\beta$, which is shown in Fig.~\ref{Fig3} for several different initial values $\beta_0$ in the absence of added rods. Our results are qualitatively consistent with recent microrheology experiments on a composite of microtubules embedded in filamentous actin \cite{kilfoil}, which reported enhanced compressibility ($\nu<0.5$) when stiff microtubules were added to an almost incompressible actin matrix ($\nu\simeq0.5$), as inferred from the measured parallel and perpendicular response functions. 

We can also calculate the effect of the addition of stiff rods by considering uniform strain of a composite, as follows. We apply a uniaxial strain $\epsilon_{zz}$ to the medium along the $z$ direction and clamp its boundaries along the $x$ and $y$ directions. We consider a particular rod making polar and azimuthal angles $\theta$ and $\phi$ with respect to the coordinate axes, as shown in the schematic Fig.~\ref{Fig1}(b). In a way similar to the above, the presence of this rigid rod is equivalent to a dipole, here of strength $p=\chi\epsilon$, where $\chi$ is the same as above, and $\epsilon=\epsilon_{zz}\cos^2\theta$. We calculate the additional stresses on the boundaries due to an isotropic distribution of such rods with number density $n$. These stresses are given by $\delta \sigma_{xx} = \delta \lambda \epsilon_{zz}$ and $\delta \sigma_{zz}= (2 \delta \mu + \delta \lambda) \epsilon_{zz}$. For rod orientations in a given solid angle $d\Omega=\sin\theta d\theta d\phi$, the stress associated with the induced dipole is given by $\delta \sigma_{ij}= \frac{n}{4\pi} \chi\epsilon  \hat{a}_i\hat{a}_jd\Omega$. Thus, the Lam\'e constants can be calculated by
 \begin{eqnarray}
 \delta \lambda &=& n\chi  \int \cos(\theta)^2  \sin(\theta)^2  \cos(\phi)^2 \frac{d\Omega}{4\pi} \nonumber \\
(2 \delta \mu + \delta \lambda)  &=&   n\chi \int \cos(\theta)^4 \frac{d\Omega}{4\pi} 
\end{eqnarray}
Solving for $\delta \mu$ and $\delta \lambda$, we find the same values as in the previous micromechanical calculation.  

We now consider the rods to have finite stretch modulus $K=\pi b^2 E_r$, where $b$ is the radius of the rod with Young's modulus $E_r$. When the rod is subject to an extensional strain $\epsilon$, the resulting force balance for an extension $\Delta$ of the rod is now given by
%
$\Delta =  a\epsilon -  \frac{2p}{\mu \pi a^2}=\frac{p}{K}$ 
%
where $p=\chi\epsilon$, and $\chi=\pi a^3\mu/(2+\pi a^2\mu/K)$. 
For a density $n$ of rods, we obtain the following nonlinear relations for the Lam\'e coefficients:
\begin{equation} 
\delta \lambda = \delta \mu =  \frac{\pi}{30} \mu n a^3/\left(1 + \frac{\pi a^2\mu}{2K}\right).\label{diffEq}
\end{equation}
In the limit of very stiff rods, this reduces to the expression above. For highly compliant rods, by contrast, this is consistent with the elastic moduli of an affinely deforming rod network of volume fraction $\varphi=\pi ab^2n$; e.g., $\delta\mu=\frac{1}{15}\varphi E_r$. 

\begin{figure}
\includegraphics[width=8cm]{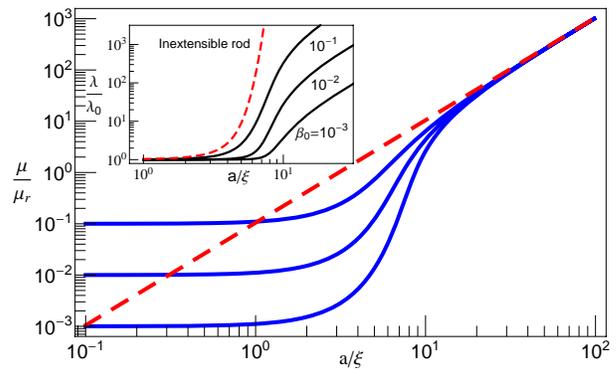}
\caption{\label{Fig4} (Color Online) The solid blue lines show the shear modulus of the composite as a function of the mesh size for different values of the ratio of the rod and medium compliance, while the dashed red line represents the affine result. The inset shows the Lam\'e coefficient $\lambda$ for different values of the initial degree of compressibility $\beta_0$ of the medium for both extensible and inextensible rods. The ratio $\mu_0/\mu_r=0.001$ in the inset, except for the inextensible rod (dashed line), where $\beta_0=0.1$.} 
\end{figure}

The result in Eq.\ (\ref{diffEq}) is valid at small density, in which the shear modulus $\mu$ on the right-hand-side is that of the (bare) matrix. If we consider $n=dn$ to be a small increase in the number density of rods, then Eq.\ (\ref{diffEq}) can be thought of as a set of differential equations representing the increase of the moduli upon the addition of stiff rods. By treating the resulting composite system as an isotropic and homogeneous effective medium, this differential equation suggests a way of calculating the properties of composites with finite rod density. This is similar to self-consistent methods employed for aligned fiber-reinforced composites \cite{Hill1965}. While this represents an uncontrolled approximation, we find that integrating Eq.\ (\ref{diffEq}) yields an exact expression for the limit of a high density of rods. The solution for $\mu$ is given by
\begin{equation}
\mu = \mu_r W \left( \frac{\mu_0}{\mu_r} \exp\left[ \frac{\mu_0}{\mu_r} + \frac{\pi n a^3}{30} \right]  \right)
\end{equation}
where,  $\mu_0$ is the shear modulus of the medium in the absence of rods and $\mu_r=2K/(\pi a^2)$. Here, $W(z)$ is the principal value of the Lambert W-function, which is defined by $z=We^W$. From this, we also obtain $\lambda=\lambda_0+\mu-\mu_0$. The results are shown in Fig.~\ref{Fig4} for various initial conditions $\mu_0$ and $\lambda_0$. For small $n$ and large $K$, this reduces to $\mu\simeq\mu_0(1+\pi na^3/30)$, since $W(z)\simeq z$ for small $z$. This is consistent with the above results for dilute, inextensible rods. As the density of rods and corresponding shear modulus increase, however, $W(z)\simeq\ln(z)$ and $\mu\rightarrow\frac{1}{15}\varphi E_r$, which is the modulus of a high density meshwork of elastic rods. 

So far, we have not accounted for the tension profile along the rod. 
Our dipole approximation is expected to overestimate the effect of the rod, since the 
real displacement field along a finite rod is expected to vary
more smoothly than for a dipole. 
The displacement field corresponds to a strain and consequent tension along the rod that is uniform at its center and vanishes at its ends. This strain field can be calculated from the force balance:
%
$Kv''(x)=\zeta\left[v(x)-\epsilon x\right]$,
%
where $v(x)$ is the displacement field along the rod in the presence of a background strain $\epsilon$ of the matrix, and $\zeta$ represents the elastic coupling of the rod to the matrix. The longitudinal strain of the rod is given by $v'(x)$, and the gradient of this corresponds to a net force per unit length on the rod, which we take to be proportional to the displacement of that section of the rod relative to the background medium. This is similar to the viscous drag on a slender body in the presence of a background velocity field \cite{batchelor}, in which case $\zeta$ is the drag coefficient per unit length of the rod. For an elastic medium, we approximate $\zeta=2\pi\mu/\ln(\xi/b)$, as in Ref.\ \cite{Cates}. Here, the screening length $\xi$ is taken to be of order the average separation or mesh size of the rod network, which depends on density. Given the weak logarithmic dependence on $\xi$, however, we will treat $\zeta$ as a constant. 

Once again, the rod can be considered as a force dipole on scales large compared with the rod length $a$. Using the condition of tension free ends, the strength of this dipole can again be expressed as $p=\chi\epsilon$, but now with 
\begin{equation}
\chi=K a \left[1-  2\ell_0\tanh\left(a/2\ell_0\right)/a\right].\label{Cates}
\end{equation} 
Here, $\ell_0=\sqrt{K/\zeta}$ represents the length over which the longitudinal state of strain of the rod varies \cite{Cates}. For highly compliant rods, this becomes a small length, corresponding to a nearly constant state of strain and tension along the rod, except very close to the ends. In the other limit, of very stiff rods, the strain exhibits a quadratic dependence, with a maximum at the center of the rod and vanishing at the ends of the rod. In this case, $\chi=\zeta a^3/12$, which is smaller by a factor of $3\ln\left(\xi/b\right)$ than the value above for the simple dipole approximation. Interestingly, independent of the parameters of the system, we still find that both Lam\'e coefficients evolve in the same way upon the addition of rods: $d\mu=d\lambda=n\chi/15$ \cite{footnote2}. This means that the qualitative form of $d\beta/dn$ in Fig.\ \ref{Fig2}, as well as our conclusions regarding the fixed points at $\beta=0$ and $\beta=1/3$, remain unchanged. 
For rods that only interact with each other through their matrix (either a viscous fluid or a homogeneous elastic matrix), the concentration of rods only enters the calculation via the screening length $\xi$. Furthermore, the modulus appearing in the coefficient $\zeta$ is that of the bare matrix. For rods that interact directly with each other, we employ a self-consistent approximation, as discussed above \cite{Hill1965}. A more detailed analysis of the effects of direct inter-fiber interactions would likely require a numerical simulation that is beyond the scope of the present work. Nevertheless, we can derive a differential equation for $\mu$ accounting for the tension profile along the rods using our self-consistent approach. Although this is more complicated than Eq.\ (\ref{diffEq}), the dipole strength derived from Eq.\ (\ref{Cates}) is well approximated by that used in Eq.\ (\ref{diffEq}), apart from the a factor of $3\ln\left(\xi/b\right)$. Thus, the functional forms in Figs.~\ref{Fig3} and \ref{Fig4} are expected to be good approximations. 

We have studied the collective mechanical response of composites of rods embedded in elastic media, such as stiff MTs in a softer cytoskeletal matrix or carbon nanotubes in synthetic gels \cite{nanotube}, using a mean field approach similar in concept to \emph{homogenization} methods introduced for elastic composites \cite{Hill1965}. We find a very general result that the addition of elastic rods or fibers leads to a monotonic evolution of Poisson's ratio toward the value $1/4$, either from above or below. On the one hand, this is consistent with recent numerical calculations for fiber-reinforced concrete, showing a weak increase in $\nu$ with fiber density in the range $0.2\le\nu<0.25$ \cite{PasaDutra}. On the other hand, our results may help to explain recent experiments \cite{kilfoil} that have reported $\nu< 1/2$ for composites of microtubules and F-actin networks, while $\nu\simeq 1/2$ for single-component F-actin networks. This suggest an important role for stiff filaments such as MTs and stress fibers in the mechanics of the cell cytoskeleton---they not only enhance the stiffness of the cytoskeleton \cite{yi-chia} and its ability to bear large forces, but may also endow it with enhanced compressibility relative to the shear compliance.  

We thank M Kilfoil, ME Cates, HNW Lekkerkerker, TC Lubensky, DC Morse and PD Olmsted for helpful discussions. MD was supported by a VENI fellowship from the Netherlands Organisation for Scientific Research (NWO). FM was supported in part by FOM/NWO.


\begin{thebibliography}{}
\bibitem{footnote2}{Although Ref.\ \cite{Cates} gives no explicit expression for $\lambda$, our results are equivalent to Eqs.\ (4.22-4.33) therein.}
\end{thebibliography}
\end{document}